\documentclass{sf2a-conf}
\usepackage{graphicx}
\begin{document}

\TitreGlobal{SF2A 2006}

\title{Be stars in open clusters in the Small Magellanic Cloud.}
\author{Martayan, C.$^{1,2}$}
\author{Baade, D.}\address{ESO, Garching-Munich, Germany}
\author{Hubert, A.-M.}\address{GEPI, Observatoire de Paris, France}
\author{Floquet, M.$^2$}
\author{Fabregat, J.}\address{Observatorio Astronomico de Valencia, Spain}
\author{Bertin, E.}\address{Institut d'Astrophysique de Paris, France}
\author{Neiner, C.$^2$}
\runningtitle{SMC Be stars.}
\setcounter{page}{1}
\index{Martayan, C.}
\index{Baade, D.}
\index{Hubert, A.-M.}
\index{Floquet, M.}
\index{Fabregat, J.}
\index{Bertin, E.}
\index{Neiner, C.}

\maketitle
\begin{abstract} 
We report on the study of the population of B and Be stars 
in SMC young clusters, performed with the Wide Field Imager 
in slitless spectroscopic mode at ESO/T2.2m with a filter
centered at H$\alpha$. First, we explain the 
reduction methods we used and our selection of different 
types of objects. Second, we present results on the proportion 
of Be stars in SMC clusters, 
and we compare this proportion to the one observed in the Milky Way.  
Finally, we also present results on a statistical study of 
variability of Be stars with OGLE.
\end{abstract}

\section{Introduction}
The main goal of these observations is to draw a map in age and metallicity of the
Be phenomenon accross the Large and Small Magellanic Clouds. 
Until now, only partial studies have given indices about the dependency of the Be phenomenon with the metallicity.
With the ESO WFI in slitless spectro. mode we obtain an exhaustive survey of the Emission line stars.
By cross-correlation with photometric databases, we obtain the photometry and classify the stars and give answers
about the origin of the Be phenomenon. 

\section{Data Reduction}
Due to the WFI mode we used, no pipeline was available to reduce the data.
We thus used the SExtractor software (Bertin \& Arnouts 1996) to 
extract the spectra (in the SMC ~1 million of spectra are used for 
this study among 3 millions available in the images). 
To recognize and distinguish emission line stars (Em*) from other 
types of stars, we created the ALBUM code. To take into account 
the defocusing across the image, this code calculates the local 
average profile of the spectra for ~50-250  isolated non Em* bright 
stars. The average profile is subtracted from each individual profile. 
In the resulting 
image, two peaks appear in the case of Em*: one is due to H$\alpha$  emission,
 the other is the residual of the filter due to the profile disturbed 
by emission. In case of non Em*, residuals of the subtraction are only
 due to the noise or non perfect shifts. We treated 86 SMC
 clusters with log(t) from 7.0 to 9 and known E[B-V] taken from the OGLE
 survey (Pietrzynski et al. 1999), which represent 7741 stars.
In addition, we obtained the astrometry with the ASTROM package 
(Wallace \& Gray 2003) with an accuracy of 0.5-1''. Then, 
we cross-correlated the catalogue of stars for each treated 
cluster to the OGLE catalogue (Szymanski 2005, Udalski et al. 1999, 
Pietrzynski et al. 1999) and we obtained the V, B and I magnitudes. 
We also used the distance modulus (Udalski et al. 2000) of the SMC 
and the calibration of Lanz (1991) to classify the stars.

\section{Be stars in open clusters in the SMC: statistics, variability}
\subsection{Proportions}
\begin{figure}[h]
  \centering
  \includegraphics[width=5cm, angle=-90]{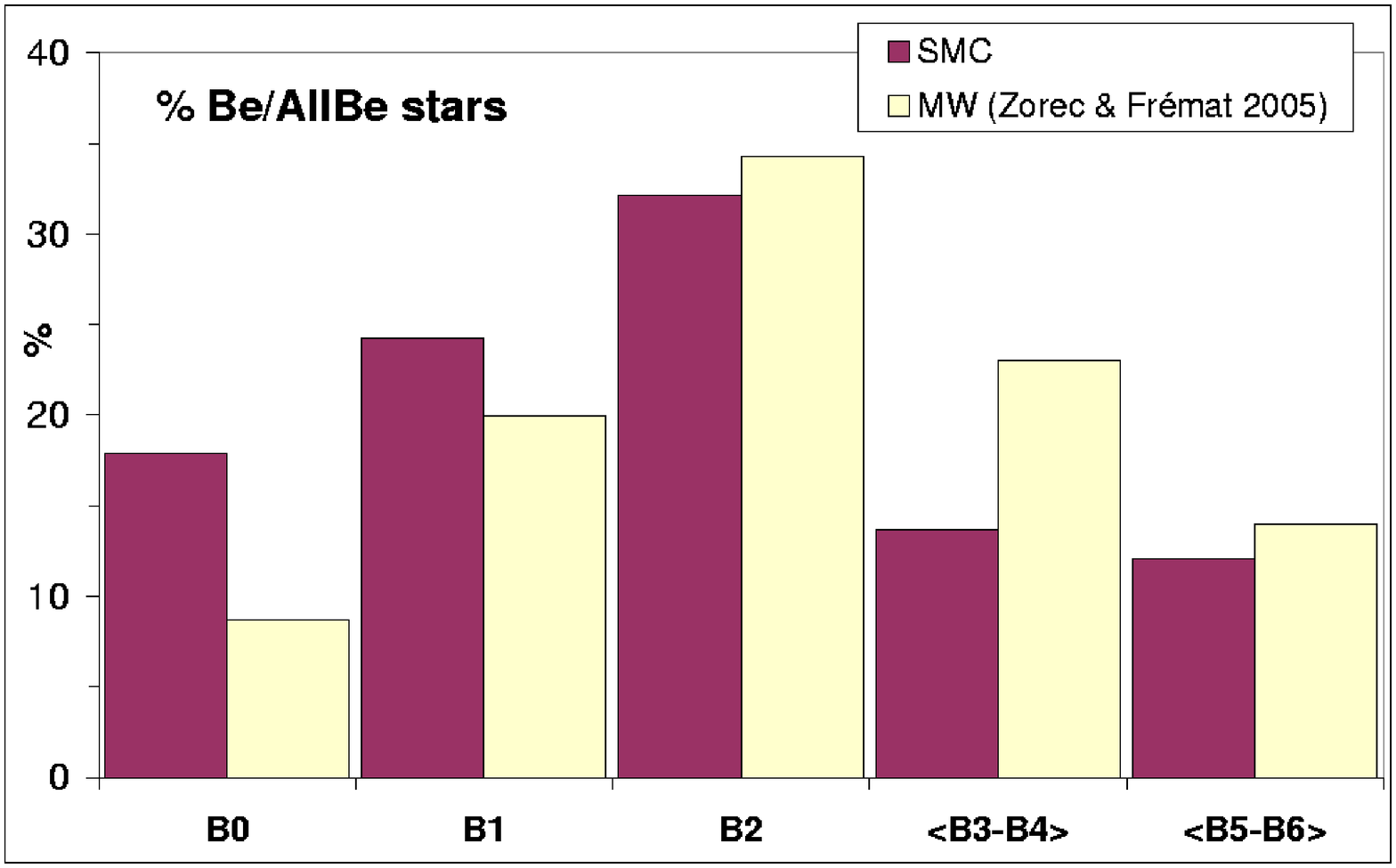}
   \includegraphics[width=5cm, angle=-90]{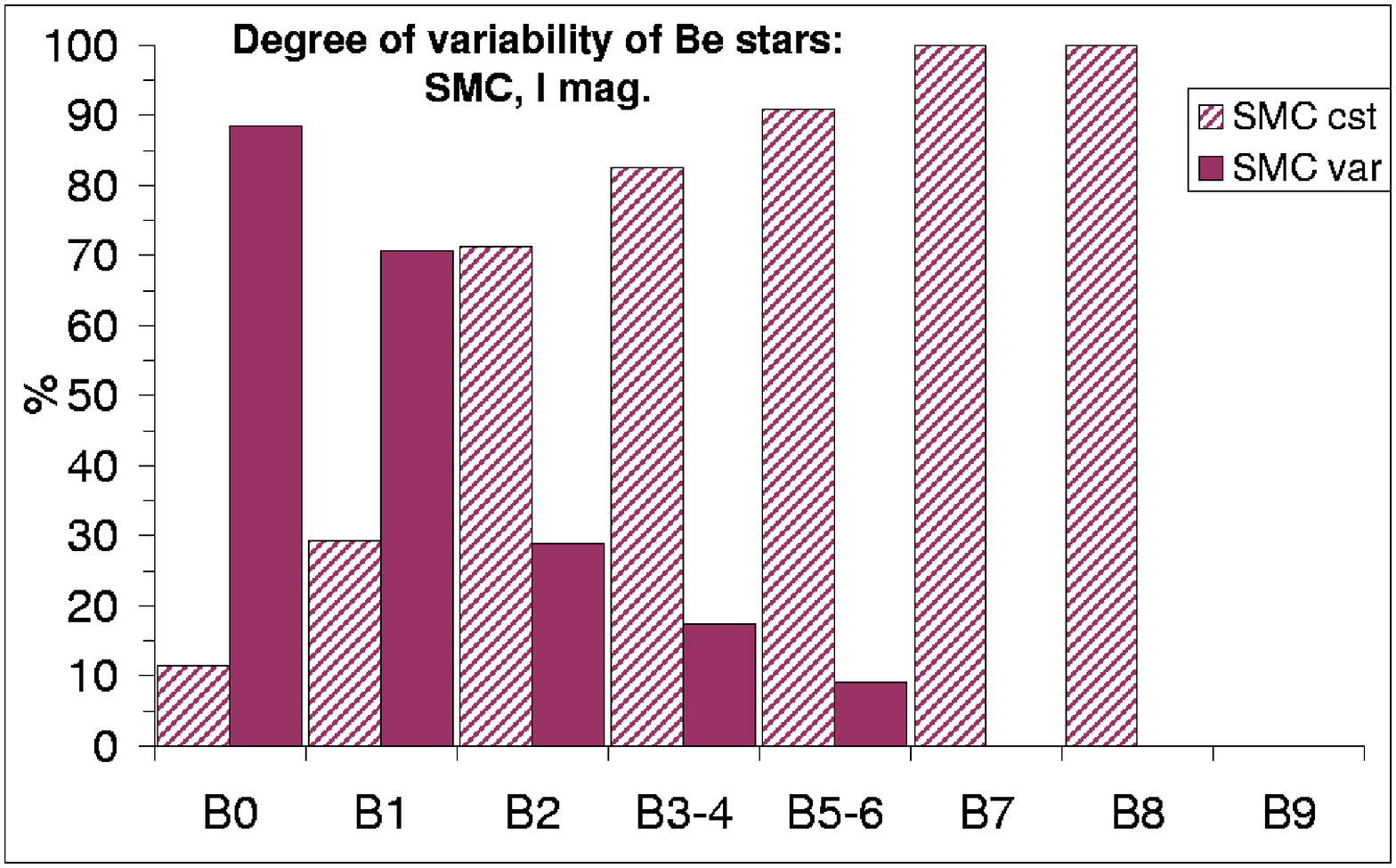}
      \caption{Left: Proportions of Be stars to the other Be stars 
      by spectral types in the SMC and in the
      MW (Zorec \& Fr\'emat 2005).
      Right: Degree of variability from the OGLE I-band 
      of candidates Be stars in the SMC open clusters.}
       \label{stats}
   \end{figure}

To study the effect of low metallicity on the Be phenomenon, 
we compare the proportions of Be stars (Be/B+Be) in open clusters 
in the SMC and MW (McSwain \& Gies 2005) by spectral types. 
We find that the lower the metallicity, the higher the rate 
of Be stars. This could be explained by higher rotational velocities in the 
SMC than in the MW (Maeder \& Meynet 2001, Martayan et al. 2006).
We also compare the ratios of Be stars by spectral types to their total 
number. Our sample ranges mainly from B0 to B6 (for late types 
there is a lack of detection due to the low S/N in their spectra). 
In the SMC as in the MW (Zorec \& Frémat 2005, Be stars in fields) the 
occurrence of Be stars in the 2 galaxies shows a maximum at B2 as shown in Fig.~\ref{stats} left.

\subsection{Variability}
We investigated the long- and short- term variability of 
Be stars in the SMC. From the OGLE database, we obtained 
light curves for 169 Be stars. 
The degree of variability of Be stars in the SMC, as shown in Fig.~\ref{stats} right, 
is higher for early types stars as in the MW (Hubert \& Floquet 1998). 

\section{Conclusions}
We obtained catalogues of Em* in 90 SMC open clusters.
We found an effect of metallicity on the proportion of Be stars. 
There are more Be stars by spectral types in the SMC than in the MW compared 
to B stars.
The distribution of Be stars versus spectral types is the same in the SMC and in the MW.
Early Be stars vary more than late-types in the SMC as in the MW.
The LMC is under investigation.



\begin{thebibliography}{}
\bibitem{}Bertin, E. \& Arnouts, S., 1996, A\&AS, 117, 393
\bibitem{}Hubert, A.-M. \& Floquet, M., 1998, A\&A, 335, 565
\bibitem{}Lanz, K. R., 1991, Eds Springer-Verlag
\bibitem{}Maeder, A. \& Meynet, G., 2001, A\&A, 373, 555
\bibitem{}Martayan, C., et al., 2006, A\&A, submitted
\bibitem{}McSwain, V. \& Gies, D., 2005, ApJS, 161, 118
\bibitem{}Pietrzynski, G., et al., 1998, Acta Astr., 48, 175
\bibitem{}Szymanski, M., 2005, Acta Astr., 55, 43
\bibitem{}Udalski, A., et al., 1999, Acta Astr., 48, 147 
\bibitem{}Udalski, A., et al., 2000, 50, 279
\bibitem{}Wallace, P. T. \& Gray, N., 2003, User Guide of ASTROM
\bibitem{}Zorec, J. \& Fr\'emat, Y., 2005, sf2a2005, 361
\end{thebibliography}
\end{document}